\documentclass[pra,twocolumn,tightenlines,showpacs,nofootinbib]{revtex4-1}
\usepackage{bm,dcolumn,amsmath,graphicx,amsfonts,amssymb}

\newcommand{\eref}[1]{Eq.~(\ref{#1})}
\newcommand{\tref}[1]{Table~\ref{#1}}

\begin{document}
\title{The atomic electric dipole moment induced by the nuclear electric dipole moment;
the magnetic moment effect}
\author{S. G. Porsev$^{1,2}$}
\author{J. S. M. Ginges$^1$}
\author{V. V. Flambaum$^1$}
\affiliation{$^1$ School of Physics, University of New South Wales,
Sydney, NSW 2052, Australia}
\affiliation{$^2$ Petersburg Nuclear Physics Institute, Gatchina,
Leningrad district, 188300, Russia}

\date{ \today }
\pacs{31.15.A-, 11.30.Er}

\begin{abstract}
We have considered a mechanism for inducing a time-reversal violating electric dipole moment (EDM)
in atoms through the interaction of a nuclear EDM $d_N$ with the hyperfine interaction, the ``magnetic moment effect''.
We have derived the operator for this interaction and presented analytical formulas for
the matrix elements between atomic states.
Induced EDMs in the diamagnetic atoms $^{129}$Xe, $^{171}$Yb, $^{199}$Hg, $^{211}$Rn, and $^{225}$Ra have been calculated numerically.
From the experimental limits on the atomic EDMs of $^{129}$Xe and $^{199}$Hg, we have placed the following constraints on the
nuclear EDMs,
$|d_N(^{129}{\rm Xe})|<  1.1 \times 10^{-21}\, |e|\,{\rm cm}$
and $|d_N(^{199}{\rm Hg})|< 2.8 \times 10^{-24}\, |e|\,{\rm cm}$.
%These limits give an accuracy benchmark for proposed measurements of nuclear EDMs in ion storage rings.

\end{abstract}

\maketitle
%=============
\section{Introduction}
%=============
Since the discovery of CP-violation in the decay of long-lived $K^0$ mesons,
there have been many efforts to observe CP or time-reversal (T) violation
in other systems. The latter is equivalent to CP-violation assuming the
validity of the CPT theorem. In particular, the possible existence of
a permanent electric dipole moment (EDM) of a particle would imply
the violation of both parity and time-reversal invariance; see, e.g., \cite{KhrLam97}.
The EDMs of the particles predicted by the standard model are too small to be detected
at the present level of experimental accuracy. However, different extensions of the standard
model (such as supersymmetry) predict much
larger EDMs of the particles that could, in principle, be found using modern experimental techniques,
see, e.g., \cite{Bar93,PosRit05}.

Recently, Griffith {\it et al.}~\cite{GriSwaLof09} reported a 7-fold improvement on the limit
of the atomic EDM of $^{199}$Hg,
$|d(^{199}{\rm Hg})| < 3.1 \times 10^{-29}\, |e|$ cm.
This is the most stringent limit on an atomic EDM.
In order to interpret the measurement in terms of fundamental P,T-violating parameters, atomic calculations
are required.
In our recent paper~\cite{DzuFlaPor09} (see also
\cite{FlaGin02,DzuFlaGin02_EDM,DzuFlaGin07})
we calculated atomic electric dipole moments induced by the nuclear Schiff moment, the P,T-odd
electron-nucleon interaction, and the electron electric dipole moment and placed limits on the
coupling constants of the P,T-odd interactions from the new Hg result.

In this work we consider one more P,T-odd interaction which gives rise to an atomic
EDM in the second order of perturbation theory.
This interaction was first discussed by Schiff in Ref.~\cite{Sch63}. He showed that if a nucleus has a permanent
EDM ${\bf d}_N$, an atomic EDM may be induced due to the interaction of ${\bf d}_N$ with the magnetic
field created by the electrons at the nucleus.

It is worth noting that the existence of a non-zero ${\bf d}_N$ alone is insufficient for producing observable
EDM effects in neutral atoms due to electronic screening of an applied electric field at the nucleus \cite{Sch63}.
This screening is exact for the case of a point-like nucleus
experiencing electrostatic forces.
One can circumvent this screening by accounting for the finite size of the nucleus or the hyperfine interaction.
The former leads to the appearance of the nuclear Schiff moment; the latter ``magnetic moment effect''
is the subject of this work.

In his work \cite{Sch63}, Schiff evaluated the EDMs of H and He induced by the magnetic moment effect.
Hinds and Sandars later calculated the effect in TlF \cite{HinSan80}. In the current work we
present a general analysis of the effect for atoms and perform calculations for diamagnetic atoms of experimental interest.

The volume effect (nuclear Schiff moment) is generally considered to be the dominant mechanism inducing EDMs in heavy
diamagnetic atoms with nuclear spin $I=1/2$. However, it has been discovered recently that the nuclear Schiff moment
is very sensitive to many-body corrections, see, e.g., \cite{BanDobEng10} and references therein.
These corrections suppress the bare values for the Schiff moments
for all considered nuclei.
Other than a general suppression, there is yet no agreement between the many-body approaches.
The nuclear EDM contribution to the atomic EDM (through the magnetic moment effect) should therefore not be
disregarded before the nuclear many-body problem is well-understood and specific CP-violating models considered.
%We hope that this work will motivate nuclear
%theorists to perform calculations for the nuclear EDM.

Further motivation for this work comes from the growing interest in measuring nuclear EDMs in ion storage rings
\cite{Khr98,FarJunMil04,OrlMorSem06,Bar08}.
In ions the nuclear EDM is not screened 
%(ion EDM = nuclear EDM $\times Z_i/Z$)
and can be measured directly.
The relations obtained in this work between the nuclear and atomic EDMs enable one to place limits on nuclear
EDMs from neutral atom measurements. These limits may be considered as an accuracy benchmark for proposed
nuclear EDM measurements with ions.

The paper is organized as follows.  In sections~\ref{Sec:GF} and \ref{Sec:EME} we derive
analytical expressions for the P,T-odd operator and the matrix elements of
this operator between atomic states. In Sec.~\ref{Sec:AEDM} we present the equation
for the atomic EDM ${\bf d}^N_{\rm at}$ and discuss different contributions to
${\bf d}^N_{\rm at}$. In Sec.~\ref{Sec:MCR} we obtain simple analytical
formulas that can be used for an estimate of the atomic EDM (and this is compared to the
atomic EDM induced by the nuclear Schiff moment), we describe our numerical
method for EDM calculations of diamagnetic atoms, and we present results for
$^{129}$Xe, $^{171}$Yb, $^{199}$Hg, $^{211}$Rn, and $^{225}$Ra.
Section~\ref{Sec:Con} contains concluding remarks.
%=================
\section{General formalism}
\label{Sec:GF}
%=================
Let us assume that the nucleus has a P,T-odd EDM,
$\mathbf{d}_{N} \equiv\langle\mathbf{d}_{N} \rangle= d_{N}\mathbf{I}/I$. Here $\langle
\mathbf{d}_{N} \rangle$ denotes the expectation value of the nuclear electric
dipole moment with the exact nuclear ground state wave function~\cite{SenAueFla08}.

The P,T-odd operator corresponding to the magnetic moment effect can be written as~\cite{Sch63}
\begin{equation}
U=-i[Q,H_{M}],
\label{Eq:U}%
\end{equation}
where square brackets in \eref{Eq:U} denote a commutator. The operator $Q$ is determined as
(if not stated otherwise we use atomic units $\hbar=m=|e|=1$)
\begin{equation}
Q=\frac{\mathbf{d}_{N} \mathbf{p}_{N}}{Z}=-\frac{1}{Z}\sum_{k=1}%
^{Z}\mathbf{d}_{N}\mathbf{p}_{e}^{(k)},
\label{Eq:Q}%
\end{equation}
where $Z$ is the nuclear charge and $\mathbf{p}_N$ and $\mathbf{p}_e^{(k)}$ are momentum operators for
the nucleus and electrons, respectively.
$H_M$ is the operator for the hyperfine interaction (HFI) which
may be interpreted as the interaction of the nuclear magnetic moment
${\bm\mu}_{N}$ with the magnetic field of the electrons.
It can be represented by the sum of the single-electron operators
\begin{equation}
H_M= \sum_{i=1}^Z \frac{\mathbf{r}_i \times {\bm\alpha}_i}{r_{i>}^{3}}{\bm\mu}_{N},
\end{equation}
where $r_{i>}\equiv\mathrm{max}(r_i,R)$ with $R$ the nuclear radius, and
$\bm\alpha=\left(
\begin{array}
[c]{cc}%
0 & \bm\sigma\\
\bm\sigma & 0
\end{array}
\right)  $ are the Dirac matrices.
${\bm \mu_N} = \mu_N {\bf I}/I$ and $\mu_N =\mu/(2\, m_p\, c)$, where $\mu$ is the nuclear magnetic moment,
$m_p$ is the nucleon mass and the speed of light $c = 1/\alpha \approx 137$.

A note on the origin of the $P,T$-odd operator Eq.~(\ref{Eq:U}). Schiff showed (we limit the explanation
to the case of an atom with nuclear EDM) that when an atom is acted
upon by electrostatic forces only and the nuclear EDM and charge distributions are
the same, the full Hamiltonian may be expressed as $H=H_0+i[Q,H_0]$, where the full Hamiltonian $H$ includes
the nuclear EDM and $H_0$ does not. The eigenvalues of $H$ are therefore the same as those of $H_0$ to first order
(the expectation value of the commutator containing $H_0$ is zero) and do not contain the nuclear EDM,
that is, there is no linear Stark shift and no observable EDM of the atom.
When the hyperfine interaction is taken into account, the full Hamiltonian may be written as
$H'=H_0'+i[Q,H_0']-i[Q,H_M]$, where the prime denotes that the magnetic interaction is included.
Only the commutator containing $H_M$ may lead to observable EDM effects.
We refer the reader to Schiff's landmark work Ref.~\cite{Sch63} for details.

In Eq.~(\ref{Eq:Q}) we assumed that the
center of mass is at rest. As a result, the momentum operator of the nucleus $\mathbf{p}_{N}$
can be replaced by the sum of the electronic momenta $\mathbf{p}_{e}^{(k)}$ with the
opposite sign. In the following we will deal with the electronic momenta only and
omit the subscript $e$, i.e., we denote $\mathbf{p}\equiv\mathbf{p}_{e}$. Then we obtain
for the operator $U$:%
\begin{equation}
U =-i \frac{d_{N}\mu_{N}}{ZI^2} \sum_{k=1}^Z
\left[  \frac{\mathbf{r}_k \times {\bm\alpha}_k }{r_{k>}^{3}}\mathbf{I},{\bf I}{\bf p}_k\right]
\equiv \sum_{k=1}^Z U_k \ .
\label{Eq:U_1}%
\end{equation}

We will consider the single-electron operator $U_k$ (omitting for brevity the index $k$) since,
as seen from \eref{Eq:U_1}, a generalization to the case of a many-electron atom is straightforward.
In addition, we restrict ourselves to consideration of the most interesting case of
nuclear spin $I=1/2$. Then for the Cartesian components $m$ and $i$ of the nuclear spin ${\bf I}$ we have
\begin{equation}
I_{m}I_{i}=\frac{1}{4}\delta_{mi}+\frac{i}{2}\varepsilon_{mil}I_{l}.
\label{Eq:Imi}
\end{equation}
Substituting \eref{Eq:Imi} in \eref{Eq:U_1} and taking into account that the terms $\sim\delta_{mi}$ will
be canceled out, after simple transformations we obtain%
\begin{equation}
U=\frac{d_{N}\mu_{N}}{Z} \left[ {\bm\alpha}{\bm \sigma}_N  \left\{
\mathbf{p},\frac{\mathbf{r}}{r_{>}^{3}}\right\}_{+} - \left\{
\frac{{\bm \sigma}_N \mathbf{r}}{r_{>}^{3}} , {\bm\alpha}%
\mathbf{p}  \right\}  _{+}\right]  . \label{Eq:U_2}%
\end{equation}
Here ${\bm \sigma}_N = 2{\bf I}$ and $\left\{  {...}\right\}  _{+}$ is an anticommutator.
Now we take into account that%
\[
\mathbf{p}=-i{\bm\nabla}=-i\mathbf{n}\frac{\partial}{\partial r}-
\frac{\mathbf{n}\times\mathbf{L}}{r} \ ,
\]
with ${\bf L} = {\bf r} \times {\bf p}$ the orbital momentum operator.

Hence%
\[
\left\{  \mathbf{p},\frac{\mathbf{r}}{r_{>}^{3}}\right\}_+=
-i\left\{ \frac{\partial}{\partial r},\frac{r}{r_{>}^{3}}\right\}_+
\]
and finally we obtain%
\begin{eqnarray}
U &=& -\frac{d_N \mu_N}{Z} \nonumber \\
&\times&\left[  i\, {\bm\alpha}{\bm \sigma}_N  \left\{
\frac{\partial}{\partial r},\frac{r}{r_{>}^{3}}\right\}_{+} +
\left\{  \frac{{\bm \sigma}_N \mathbf{r}}{r_{>}^{3}}, {\bm\alpha}%
\mathbf{p}  \right\}_{+}\right] .
\label{Eq:U_we}
\end{eqnarray}

If the typical distances of interest are small one can neglect the eigenvalue and the mass of the electron
in comparison with the electrostatic potential. Hinds and Sandars showed in Ref.~\cite{HinSan80}
that in this approximation the operator $U$ can be written in a more simple form:
\begin{equation}
U = \frac{2\, d_N \mu_N}{Z}\, {\bm \sigma}_N \frac{{\bm \alpha} \times {\bf L}}{r^3} .
\label{Eq:U_HS}
\end{equation}
In the following sections we will discuss the difference between the two forms of the operator
$U$ given by  Eqs. (\ref{Eq:U_we}) and (\ref{Eq:U_HS}).
%===========================================
\section{Electronic matrix elements}
\label{Sec:EME}
%===========================================
We use the following form for the electronic wave functions%
\[
|n\varkappa m\rangle=\left(
\begin{array}
[c]{c}%
f(r)\Omega_{jlm}\\
ig(r)\Omega_{j\widetilde{l}m}%
\end{array}
\right)  =\left(
\begin{array}
[c]{c}%
f(r)\Omega_{\varkappa m}\\
ig(r)\Omega_{-\varkappa m}%
\end{array}
\right)  ,
\]
where $\tilde{l}\equiv2j-l$.
Using the  expression for the operator $U$ given by \eref{Eq:U_we} we can
derive the electronic matrix elements (MEs).
Finally we find (see the Appendix for details of the derivation)%
\begin{align}
&\left\langle n^{\prime}\varkappa^{\prime}m^{\prime}\left\vert U\right\vert
n\varkappa m\right\rangle =-\frac{d_N \mu_N}{Z}\, {\bm \sigma}\!_N \!\left\langle
\varkappa^{\prime}m^{\prime}\left\vert \mathbf{n}\right\vert \varkappa
m\right\rangle \nonumber\\
&  \times\int_{0}^{\infty}\left[  \frac{\varkappa+\varkappa^{\prime}}%
{r}\left\{  f^{\prime}g\left(  \varkappa-\varkappa^{\prime}+1\right)
+fg^{\prime}\left(  \varkappa^{\prime}-\varkappa+1\right)  \right\} \right.
\nonumber\\
&  +\left.
 \left(  \varepsilon^{\prime}-\varepsilon \right)  \left(
\varkappa - \varkappa^{\prime} \right)  \left(  f^{\prime}f-g^{\prime}g\right)
\right]  \frac{r^{3}}{r_{>}^{3}}dr.
\label{Eq:U_f}%
\end{align}

If we factorize $(\varepsilon' -\varepsilon)$ in the second term of \eref{Eq:U_f},
we can rewrite it as
\begin{equation}
(\varepsilon -\varepsilon') \frac{d_N \mu_N}{Z}\, {\bm \sigma}\!_N
\left\langle n' \varkappa' m'
\left\vert \, i\, \frac{{\bf r} \times {\bm \sigma}}{r_>^3} \right\vert n \varkappa m \right\rangle.
\end{equation}
Then \eref{Eq:U_f} can be represented by
\begin{align}
&\left\langle n^{\prime}\varkappa^{\prime}m^{\prime}\left\vert U\right\vert
n\varkappa m\right\rangle =-\frac{d_N \mu_N}{Z} {\bm \sigma}\!_N  \! \left[ \left\langle
\varkappa^{\prime}m^{\prime}\left\vert \mathbf{n}\right\vert \varkappa
m\right\rangle \right. \times \nonumber\\
&  \left.  (\varkappa+\varkappa^{\prime})\int_{0}^{\infty}%
\left\{  f^{\prime}g\left(  \varkappa-\varkappa^{\prime}+1\right)
+fg^{\prime}\left(  \varkappa^{\prime}-\varkappa+1\right)  \right\}  \frac{r^{2}dr }{r_{>}^{3}}\right.
\nonumber\\
&  +\left. (\varepsilon' -\varepsilon) \left\langle n' \varkappa' m'
\left\vert\, i \frac{ {\bf r} \times {\bm \sigma}}{r_>^3} \right\vert n \varkappa m \right\rangle  \right] .
\label{Eq:U_f1}%
\end{align}

As seen from Eqs. (\ref{Eq:U_f}) and (\ref{Eq:U_f1}), the first term in these equations disappears if
$\varkappa'+\varkappa = 0$. It happens for the states with $l' = l \pm 1$ and $j'=j$.
As a result, the MEs $\left\langle -\varkappa m' \left\vert
U\right\vert \varkappa m \right\rangle$  turn out to be proportional  to $\left(  \varepsilon' -\varepsilon \right)$.
For heavy atoms these MEs are small. In
particular, it means that the MEs $\left\langle s\left\vert U\right\vert p_{1/2}\right\rangle $
contribute less to the atomic EDM than the MEs $\left\langle s\left\vert
U\right\vert p_{3/2}\right\rangle $.
If we neglect the term $\sim\left(  \varepsilon^{\prime} -\varepsilon\right)$ in  Eqs. (\ref{Eq:U_f}) and (\ref{Eq:U_f1}),
then Eqs. (\ref{Eq:U_we}) and (\ref{Eq:U_HS}) lead to the same formula for the MEs.
%====================
\section{Atomic EDM}
\label{Sec:AEDM}
%====================
In this section we present our calculations of EDMs induced by the operator $U$ for
several diamagnetic atoms of experimental interest having $I=1/2$.
The EDM $\mathbf{d}_{\rm at}^N = d_{\rm at}^N\, {\bm \sigma}\!_N$
of an atom in the state $|0 \rangle$ in the second order of perturbation theory is given by
\begin{equation}
\mathbf{d}_{\mathrm{at}}^N =2\sum_K
\frac{\left\langle 0\left\vert \mathbf{D}\right\vert K\right\rangle
\left\langle K\left\vert U\right\vert 0\right\rangle }{E_{0}-E_{K}},
\label{Eq:Dat0}
\end{equation}
where ${\bf D}=-{\bf r}$ is the electric dipole operator and the summation goes over
all intermediate states $|K \rangle$ allowed by the selection rules.

As follows from the consideration given in the preceding section, it is convenient to present
the operator $U$ as a product of two operators, one relating to the electronic part and
the other relating to the nuclear part, i.e.
\begin{equation}
U={\bf U}_{\rm{el}}\, {\bm \sigma}\!_N,
\label{Eq:UelI}
\end{equation}
where $\mathbf{U}_{\mathrm{el}}$ describes the electronic part of the operator $U$.

Applying the Wigner-Eckart theorem, and summing over magnetic
quantum numbers $m$ of the initial and final states in~\eref{Eq:Dat0},
we obtain for $d^N_{\mathrm{at}}$:%
\begin{equation}
d_{\mathrm{at}}^N=\frac{2}{3}\sum_K
\frac{\left\langle 0\left\vert | r \right\vert |K\right\rangle
\left\langle 0\left\vert |U_{\mathrm{el}}\right\vert |K\right\rangle }{E_{K}-E_{0}}.
\label{Eq:Dat}%
\end{equation}

To carry out calculations of atomic EDMs for atoms with closed shells, it is
convenient to rewrite Eqs.~(\ref{Eq:Dat0}) and (\ref{Eq:Dat}) in terms of single-electron wave
functions. Then the corresponding expressions for $\mathbf{d}_{\mathrm{at}}^N$ and
$d_{\rm{at}}^N$ are given by%
\begin{equation}
\mathbf{d}_{\mathrm{at}}^N =2\sum_{c,n}
\frac{\left\langle c \left\vert \mathbf{r}\right\vert n \right\rangle
\left\langle n\left\vert U\right\vert c\right\rangle }{\varepsilon_{n}-\varepsilon_{c}}
\label{Eq:Dats0}
\end{equation}
and
\begin{equation}
d_{\mathrm{at}}^N=\frac{2}{3} \sum_{c,n}
\frac{\left\langle n ||r|| c\right\rangle \left\langle
n\left\vert |U_{\mathrm{el}}\right\vert |c\right\rangle }{\varepsilon
_{n}-\varepsilon_{c}},
\label{Eq:Dats}%
\end{equation}
where the indices $c$ and $n$ relate to the single-electron
core and virtual orbitals and $\varepsilon_{c(n)}$ are the single-electron core (virtual) energies, correspondingly.

Note that if we substitute the second term of \eref{Eq:U_f1} into \eref{Eq:Dats0}, then after cancelation of
$({\varepsilon_n-\varepsilon_c})$ in the numerator and denominator and applying closure
$\sum_n | n \rangle \langle n | =1$, we obtain a term which is proportional to
\begin{equation*}
\sum_c  \left\langle c \left\vert\, \frac{{\bf r}\, [({\bf r} \times {\bm \sigma})\, {\bm \sigma}_N]}{r_>^3}
\, \right\vert c \right\rangle.
\label{Eq:2nd}
\end{equation*}
It can be readily shown that if we apply again the Wigner-Eckart theorem and sum up over the projections $m_c$ of the total
angular momenta $j_c$, this term goes to zero and therefore does not contribute to
${\bf d}_{\rm at}^N$. In particular, it means that the second term in \eref{Eq:U_f1} does not contribute
to the atomic EDM for closed-shell atoms.

Keeping in mind that we intend to carry out calculations of EDMs for atoms with closed shells, we can neglect the term
$\sim (\varepsilon' - \varepsilon)$ in Eq.~(\ref{Eq:U_f1}). Accounting for Eq.~(\ref{Eq:UelI}), we obtain the following
expression for the reduced ME of the operator $U_{\mathrm{el}}$:
\begin{align}
\left\langle n^{\prime}\varkappa^{\prime}|\left\vert U_{\mathrm{el}%
}\right\vert |n\varkappa\right\rangle &\approx -\frac{d_{N}\mu_{N}}{Z}%
\left\langle \varkappa^{\prime}|\left\vert n |\right\vert \varkappa
\right\rangle (\varkappa+\varkappa') \nonumber \\
 \times \int_{0}^{\infty}
&\left\{  f^{\prime}g\left(  \varkappa-\varkappa^{\prime}+1\right) \right. \nonumber \\
&\left. + fg^{\prime}\left(  \varkappa^{\prime}-\varkappa+1\right)  \right\}  \frac{dr}{r},
\label{Eq:Uel_red}
\end{align}
where the reduced ME
$\left\langle \varkappa' ||n|| \varkappa \right\rangle $
is given by%
\begin{align}
\left\langle \varkappa^{\prime}|\left\vert n |\right\vert \varkappa
\right\rangle & =\left(  -1\right)  ^{j^{\prime}+1/2}\sqrt{\left(  2j^{\prime
}+1\right)  \left(  2j+1\right)  } \nonumber \\
& \times \left(
\begin{array}
[c]{ccc}%
j^{\prime} & j & 1\\
-1/2 & 1/2 & 0
\end{array}
\right)  \xi\left(  l^{\prime}+l+1\right)
\label{Eq:n}
\end{align}
with%
\[
\xi(x)  =\left\{
\begin{array}
[c]{c}%
1,\text{ if }x\text{ is even}\\
0,\text{ if }x\text{ is odd}%
\end{array}
\right. .
\]

In \eref{Eq:Uel_red} we also took into account that the integral inside the nucleus is very small and replaced $r_>$ by $r$.
As we mentioned above, the MEs of the operator
$U_{\rm el}$ turn to zero if $\varkappa+\varkappa' = 0$. It means that MEs
between states with the same total angular momentum though different parity like $\langle s || U_{\rm el} || p_{1/2} \rangle$,
$\langle p_{3/2} || U_{\rm el} || d_{3/2}\rangle$, etc. are equal to zero.
%=============================================
\section{Method of calculation and results}
\label{Sec:MCR}
%=============================================
%=============================================
\subsection{Analytical estimates}
\label{SubSec:AE}
%=============================================
In this section we derive the analytical expression for the
matrix element ~\eref{Eq:Uel_red}.
Outside the nucleus the wave functions $f_{n \varkappa}$
and $g_{n \varkappa}$ can be represented by~\cite{Khr91}%
\begin{align}
&f_{n \varkappa} = \frac{\varkappa}{|\varkappa|} \frac{1}{{\sqrt{Z\nu^3}}r}
\left[  (\gamma+\varkappa)J_{2\gamma}(x) - \frac{x}{2}J_{2\gamma-1}(x) \right], \nonumber \\
&g_{n \varkappa}=\frac{\varkappa}{|\varkappa|} \frac{1}{{\sqrt{Z\nu^3}}r} Z \alpha\, J_{2\gamma}(x) ,
\end{align}
where $x \equiv \sqrt{8Zr}$, $\gamma=\sqrt{\varkappa^{2}-Z^{2}\alpha^{2}}$, and
$J_\nu(x)$ are Bessel functions.

Using these expressions for the radial wave functions we obtain for the radial integral
\begin{equation}
\int_{0}^{\infty}f_{n^{\prime}\varkappa^{\prime}}g_{n \varkappa}\frac{dr}{r}=
\frac{Z^{2}\alpha}{\left(  \nu^{\prime}\nu\right)  ^{3/2}} \lambda R_{M}.
\label{Eq:int}%
\end{equation}
Here
\begin{eqnarray}
\lambda \equiv \frac{\varkappa^{\prime}\varkappa}{|\varkappa^{\prime}\varkappa|} 96
\left[  \varkappa'-1 + \frac{1}{4}
\left(  \varkappa^2 - \varkappa'^{2} \right)  \right]  A_{j'j} ,
\label{Eq:lambda}%
\end{eqnarray}
the relativistic enhancement factor $R_{M}$ is given by~\cite{SusFlaKhr84}%
\begin{equation*}
R_{M}\equiv\frac{1}{A_{j'j}} \frac{\Gamma(\gamma^{\prime}+\gamma-2)}
{\Gamma(\gamma^{\prime}-\gamma+3) \Gamma(\gamma- \gamma^{\prime}+3)
\Gamma(\gamma^{\prime}+ \gamma+3)} ,
\label{Eq:RM}%
\end{equation*}
$\Gamma(\beta)$ are the $\gamma$-functions, and the factor $A_{j'j}$ is determined as follows
\begin{equation}
A_{j'j} \equiv\frac{(j^{\prime}+j-2)!}{(j^{\prime}-j+2)! (j-j^{\prime
}+2)!(j^{\prime}+j+3)!}.
\end{equation}

Let us consider an important particular case of the ME $\langle s ||U_{\rm el}||p_{3/2} \rangle$.
In this case $\varkappa'=-1$ and $\varkappa=-2$. Then the first term under the integral
in~\eref{Eq:Uel_red} turns to zero while for the second term in the integral we find,
using Eqs. (\ref{Eq:int}) and (\ref{Eq:lambda}),
\begin{equation}
\int_{0}^{\infty}f_{n p_{3/2}}g_{n^{\prime}s}\frac{dr}{r} = -\frac{1}{2}%
\frac{Z^{2}\alpha} {\left(  \nu'_s \nu_{p_{3/2}}\right)
^{3/2}}R_{M}.
\end{equation}
For the matrix element we therefore have
\begin{equation}
\langle n^{\prime}s || U_{\rm el} ||  np_{3/2} \rangle \approx
d_N \mu_N \frac{2\sqrt{3}\, Z \alpha}{\left(  \nu'_s \nu_{p_{3/2}} \right) ^{3/2}}R_M.
\end{equation}
This analytical expression demonstrates how the atomic EDM depends on $Z$ and nuclear parameters.
The enhancement factors $R_M$ for the medium atoms are close to unity. They grow with increasing $Z$ approaching the
value 2.2 for Ra. The values of the enhancement factors for the diamagnetic atoms considered in this work are listed in \tref{T:RM}.

A simple analytical estimate gives the following result for the atomic EDM
\begin{equation}
d_{\rm at}^N \approx \pm \,10^{-7}\mu ZR_M d_N ,
\end{equation}
where the upper sign (+) relates to the divalent atoms and the lower
sign (-) relates to the noble gases.
This formula gives values in
reasonable  agreement  (within a factor of $\sim 2$) with the many-body 
Dirac-Hartree-Fock (DHF) results presented in
the following section.

It is instructive to compare this estimate with a similar one
for the atomic EDM induced by the nuclear Schiff moment $S$,
\begin{equation}
d_{\rm at}^S\approx \mp \, 10^{-22}Z^2(R_{1/2}+2R_{3/2})~S/(|e|~{\rm fm}^3)~|e|~{\rm cm} \ ,
\end{equation}
where $R_{1/2}$ and $R_{3/2}$ are relativistic enhancement factors corresponding
to $s-p_{1/2}$ and $s-p_{3/2}$ weak matrix elements, respectively; see Ref. \cite{SusFlaKhr84}
for these factors.
Again, the upper sign is for the divalent atoms and the lower sign is for noble gases.
The values obtained from this simple formula also
reasonably agree (within a factor of 2) with the many-body DHF results obtained
in Refs.~\cite{DzuFlaGin02_EDM,DzuFlaGin07,DzuFlaPor09}.
%Note that in Ref.~\cite{DzuFlaPor09} there is
%a typo in Eq.(7) where the operator of the nuclear Schiff moment was determined.
%The sign in the right hand sign of Eq.(7) should be opposite.

The atomic EDM induced by the Schiff moment
benefits from an extra $Z$ dependence which becomes very important in heavy atoms \cite{Khr91}.
In Sec.\ref{SubSec:coupling} we consider a specific mechanism for inducing the Schiff and
nuclear dipole moments and compare the sizes of the induced atomic EDMs.

%########################################################################
\begin{table}%[bt]
\caption{The nuclear charges $Z$ and the relativistic enhancement factors $R_M$.}

\label{T:RM}

\begin{ruledtabular}
\begin{tabular}{lccccc}
        & $^{129}$Xe & $^{171}$Yb & $^{199}$Hg & $^{211}$Rn & $^{225}$Ra \\
\hline
   $Z$ &       54    &   70       &   80       &    86      &     88         \\
 $R_M$ &     1.28    &   1.56     &  1.83      &  2.06      &    2.15
\end{tabular}
\end{ruledtabular}
\end{table}
%########################################################################

%-----------------------------------------------------------------
\subsection{Numerical method of calculation and results}
\label{SubSec:method}
%-----------------------------------------------------------------
Here we describe the simple numerical methods we use for calculations of atomic EDMs
for closed-shell atoms. At the first stage we solve
DHF equations in the $V^N$ approximation. This means that we include
all electrons forming the ground state of the atom in a self-consistency procedure
\begin{equation}
 H_0\, \psi_c = \varepsilon_c \,\psi_c .
\end{equation}
Here $H_0$ is the relativistic Hartree-Fock Hamiltonian and
$\psi_c$ are single-electron wave functions of the core.

To take into account polarization of the atomic core by external fields (the electric dipole field or the
P,T-odd field), we solve the random phase approximation (RPA) equations
\begin{equation}
(H_0 -\varepsilon_c)\delta \psi_c = -(F + \delta V^N) \psi_c ,
\label{Eq:RPA}
\end{equation}
where $F$ is the operator of the external field and $\delta V^N$ is the correction to the
self-consistent potential due to the effect of the external field. The RPA equations (\ref{Eq:RPA}) are solved
self-consistently for all states in the core.

In implementing the DHF and RPA procedures for calculations of the atomic EDMs, we have used two
equivalent approaches. The first involves the construction of virtual orbitals and summation over states.
The second involves the direct solution of the perturbed orbital $\delta \psi_c$ on the grid.

In the former method the virtual orbitals are constructed by multiplication of the
previous orbital of the same partial wave to a smooth function of $r$
with subsequent orthogonalization of this orbital to the rest of the orbitals.
This method was described in detail in Refs.~\cite{KozPorFla96} and~\cite{Bog91}.

In the latter method, implemented in Ref.~\cite{DzuFlaGin02_EDM} for calculation of the Schiff moments,
instead of direct summation over virtual states in Eq.~(\ref{Eq:Dats0}), we evaluate
${\bf d}_{\rm at}^N=2\sum_c \langle c|-{\bf r}|\delta c ^{U}\rangle$,
where $|\delta c^{U}\rangle =\sum_n \frac{\langle n|U|c\rangle}{\epsilon_c -\epsilon_n}|n\rangle$ is a solution of
Eq.~(\ref{Eq:RPA}) for the P,T-odd field.
In taking into account core polarization by the fields, the correction goes to one field and not the other
to avoid double counting. See Ref.~\cite{DzuFlaGin02_EDM} for details.

The results of numerical calculations carried out using the DHF and RPA methods for the considered
diamagnetic atoms are presented in~\tref{T:Dat}.
%########################################################################
\begin{table}%[bt]
\caption{The values of $d_{\rm at}^N$ in units ($10^{-6}\, d_N$) obtained in the DHF and RPA approximations.}

\label{T:Dat}

\begin{ruledtabular}
\begin{tabular}{lccccc}
& $^{129}\rm{Xe}_{54}$ & $^{171}\mathrm{Yb}_{70}$ & $^{199}\mathrm{Hg}%
_{80}$ & $^{211}\mathrm{Rn}_{86}$ & $^{225}\mathrm{Ra}_{88}$ \\
\hline
 DHF  &  4.4  &  2.6  &  3.2  &  -9.1  &  -9.5 \\
 RPA  &  5.8  &  11  &  11   &  -13   &  -33
\end{tabular}
\end{ruledtabular}
\end{table}
%########################################################################
It is seen that inclusion of the RPA corrections increases the size of the
atomic EDM. For the noble gases (Xe and Rn) the RPA corrections contribute at the level of
30--40\%, while for atomic Hg, Yb, and Ra, which have two $s$ electrons above closed
shells, the RPA corrections are much larger. In fact, they increase the EDMs
of these atoms several times compared to the DHF values. The reason for this increase is
that the two $s$ electrons are loosely bound and can be easily excited. As a result, account of
higher orders of perturbation theory (like the RPA corrections) leads
to a significant change in the ``bare'' results obtained in the DHF approximation.
This result is similar to that obtained for other contributions to the P,T-odd atomic EDM
discussed in Refs.~\cite{DzuFlaGin02_EDM,DzuFlaGin07,DzuFlaPor09} for the same atoms.

We note that in our previous works \cite{DzuFlaGin02_EDM,DzuFlaPor09}, a check on the
RPA results for Yb, Hg, and Ra was carried out by performing more sophisticated calculations in the
$V^{N-2}$ approximation. In this approach, correlations between the two valence electrons and the core
were taken into account using many-body perturbation theory (MBPT) while correlations between the
valence electrons were accounted for using the configuration
interaction (CI) method; see Refs.~\cite{DzuFlaKoz96b,DzuKozPor98} regarding the CI+MBPT method.
It was found in~\cite{DzuFlaGin02_EDM,DzuFlaPor09} that the two approaches ($V^{N}$ and $V^{N-2}$) yield
results for the EDMs that differ by less than $20\%$ for the various P,T-odd mechanisms. The
good agreement of the results obtained using two very different approaches is a strong argument in favour
of the stability of the RPA results.
Because the operator considered in this work is of similar form to the other $P,T$-odd operators, we expect
that our RPA results for the atomic EDMs are accurate to about 20\%.

Using the experimental limits on the P,T-odd atomic electric dipole moments of $^{129}$Xe~\cite{RosChu01}
\begin{align}
&d(^{129}{\rm Xe}) = (0.7 \pm 3.3_{\rm stat} \pm 0.1_{\rm syst})
                                   \times 10^{-27}\, e\,{\rm cm} \nonumber \\
&\longrightarrow
|d(^{129}{\rm Xe})| < 6.6 \times 10^{-27}\, |e|\,{\rm cm}
\label{Eq:Xe}
\end{align}
 and $^{199}$Hg~\cite{GriSwaLof09}
\begin{align}
&d(^{199}{\rm Hg}) = (0.49 \pm 1.29_{\rm stat} \pm 0.76_{\rm syst})
                                   \times 10^{-29}\, e\,{\rm cm} \nonumber \\
&\longrightarrow
|d(^{199}{\rm Hg})| < 3.1 \times 10^{-29}\, |e|\,{\rm cm}
\label{Eq:Hg}
\end{align}
we are able to place constraints on the nuclear EDMs $d_N(^{129}{\rm Xe})$ and
$d_N(^{199}{\rm Hg})$. Using Eqs.~(\ref{Eq:Xe}), (\ref{Eq:Hg}),
and the results presented in \tref{T:Dat} we obtain
\begin{align}
& |d_N(^{129}{\rm Xe})| <  1.1 \times 10^{-21}\, |e|\,{\rm cm}, \nonumber \\
& |d_N(^{199}{\rm Hg})| < 2.8 \times 10^{-24}\, |e|\,{\rm cm}.
\end{align}
We have not included the atomic theory error in these limits.
%-----------------------------------------------------------------
\subsection{Contributions to the nuclear EDM}
\label{SubSec:coupling}
%-----------------------------------------------------------------
For spherical nuclei with spin determined by a single unpaired nucleon,
there are several main terms that contribute to the nuclear EDM. One of them is
characterized by the P,T-odd nucleon-nucleon interaction while the other ones are the contributions
from the EDMs of the neutron ($d_n$) and the proton ($d_p$). 
Thus, we can write $d_N$ as~\cite{SusFlaKhr84}
\begin{align}
d_N =  d^\eta_N + t_I d_n + a_p d_p,
\label{Eq:d_N1}
\end{align}
where we denote by $d^\eta_N$ the contribution from the P,T-odd nucleon-nucleon interaction;
$ t_I =1$ for $I=l_{I}+1/2$ and $t_I =-I/(I+1)$ for $I=l_{I}-1/2$ (with $l_I$ being the orbital momentum
of the unpaired nucleon) and the coefficient $a_p$ is numerically close to 0.1.

In Refs.~\cite{SusFlaKhr84,FlaKhrSus86} it was shown that the nuclear EDM induced by the P,T-odd nucleon-nucleon interaction
can exceed the nucleon EDM by more than two orders of magnitude. If we neglect the terms
proportional to $d_n$ and $d_p$ in \eref{Eq:d_N1} and express
$d^\eta_N$ through the T-odd nucleon-nucleon coupling constant $\eta$, we obtain~\cite{FlaKhrSus86}
\begin{equation}
d_{N} \approx  4\times10^{-13}\left(  q-\frac{Z}{A}\right)  t_I \eta \ ,
\end{equation}
where $q=0$ and $1$ for an outer neutron and
proton, respectively.  Taking into account that for all atoms considered $q=0$,
we obtain for $d_N$ (in units of $|e|\, {\rm cm}$)
\begin{align}
d_N  \approx -2 \times 10^{-21} \frac{Z}{A}\, t_I \eta\, |e|\, {\rm cm} .
\label{Eq:d_N2}
\end{align}
Interactions of the outer neutron with both protons and neutrons of the core contribute to the nuclear EDM;
the coupling constant $\eta=\frac{Z}{A}\eta_{\rm np}+\frac{N}{A}\eta_{\rm nn}$.
Using \eref{Eq:d_N2} and the results for $d^N_{\rm at}$ given in \tref{T:Dat} we can express the values of the atomic EDMs through $\eta$.
These results for $^{129}$Xe, $^{171}$Yb, $^{199}$Hg, and $^{211}$Rn are presented in \tref{T:nucl}.

To get an idea of the relative size of the induced atomic EDMs compared to those induced by the Schiff moment, we consider
the Schiff moment produced by the nucleon-nucleon interaction. For all considered atoms, the nuclei have an outer neutron.
This means that there is no direct contribution to the Schiff moment. The protons need to be excited to distort the charge
density and create a P,T-odd charge distribution. A simple numerical calculation in the Woods-Saxon potential with spin-orbit
interaction included gives for $^{199}$Hg, $S(^{199}{\rm Hg})=-1.4\times 10^{-8}\eta_{\rm np}|e|~{\rm fm}^3$ \cite{FlaKhrSus86}.
Using this result, the calculation for the atomic EDM induced by the Schiff moment
$d_{\rm at}^{S}=-2.8\times 10^{-17}~S/(|e|~{\rm fm}^3)~|e|~{\rm cm}$ \cite{DzuFlaGin02_EDM,DzuFlaPor09}, and
the result for the atomic EDM induced by $d_N$ through $\eta$ presented in \tref{T:nucl}, we obtain
\begin{equation}
\left\vert \frac{d_{\rm at}^N(^{199}{\rm Hg})}{d_{\rm at}^{S}(^{199}{\rm Hg})}\right\vert \approx
0.01 \frac{(0.4\eta_{\rm np}+0.6\eta_{\rm nn})}{\eta_{\rm np}} .
\end{equation}
(The ratio is larger for lighter atoms.)
While this indicates that the contribution to the atomic EDM from the nuclear EDM is significantly smaller than that from the
nuclear Schiff moment, we remind the reader that we have used a very simple model for the nucleus.
Indeed, it is only recently that many-body calculations have been performed for the nuclear Schiff moment and these
have demonstrated that many-body corrections are large and lead to a suppression of the bare results for all nuclei
considered, see, e.g., the most recent calculation~\cite{BanDobEng10} and references therein.
The results of the different many-body approaches, however, are not in agreement.
For example, in their RPA approach, Dmitriev and Sen'kov~\cite{DmiSen03} find a suppression of two orders of magnitude in
the isoscalar channel of the P,T-odd pion-nucleon-nucleon interaction and a suppression of one order of magnitude in the isotensor
channel for the case of $^{199}$Hg.
In the fully self-consistent approach of Ban {\it et al.}~\cite{BanDobEng10} applied to $^{199}$Hg, one order of magnitude suppression
is seen in the isoscalar
and isotensor channels, while instabilities are seen in the isovector channel, with results even varying in sign.
Therefore, until the many-body problem is well-understood, and specific CP-violation models considered,
the contribution to the atomic EDM from the nuclear dipole moment should not be dismissed.

It is worth noting that there are nuclei with octupole deformation, such as, e.g., $^{223}$Rn and $^{223,225}$Ra.  For these nuclei,
the nuclear EDM cannot be approximated by the simple formula \eref{Eq:d_N2}. As shown in \cite{AueFlaSpe96,SpeAueFla97}, the P,T-odd
nuclear forces lead to an enhanced collective dipole moment that can significantly exceed single-particle moments. The enhancement of
the nuclear EDM is also possible in nuclei with quadrupole deformations due to mixing of close opposite-parity levels
\cite{HaxHen83,SusFlaKhr84}.
%#########################################################################
\begin{table}%[bt]
\caption{$I^p$ is the spin and parity of the nuclear ground state and
$\mu$ is the magnetic moment expressed in nuclear magnetons~\cite{webel}.
The values of $d_{\rm at}^N$ in units ($10^{-27}\,\eta\, |e|\, {\rm cm}$)
are obtained in the RPA approximation.}

\label{T:nucl}

\begin{ruledtabular}
\begin{tabular}{lcccc}
                      &    $I^p$  &  $\mu$  &  $t_I$  &  $d_{\rm at}^N$ \\
\hline
$^{129}$Xe  & 1/2$^+$ & -0.7780 &    1       &   -5.1 \\
$^{171}$Yb  & 1/2$^-$  &  0.4919 &   -1/3    &   3.1 \\
$^{199}$Hg  & 1/2$^-$  &  0.5059 &  -1/3    &    3.1  \\
$^{211}$Rn  & 1/2$^-$  &  0.60     &  -1/3   &   -3.7  \\
$^{225}$Ra  & 1/2$^+$ & -0.734   &    1      &  \\ % 27   \\
\end{tabular}
\end{ruledtabular}
\end{table}
%########################################################################

%=============================================
\section{Conclusion}
\label{Sec:Con}
%=============================================
We have derived an expression for the P,T-odd operator produced by the interaction
of a P,T-odd electric dipole moment of the nucleus with the operator of the
hyperfine interaction. We have presented simple analytical
formulas that can be used for an estimate of the EDM for different atoms. Using numerical
methods, we have found the contributions to atomic EDMs caused by this P,T-odd
interaction for a number of diamagnetic atoms. Using the experimental limits
on the atomic electric dipole moments of $^{129}$Xe and $^{199}$Hg, we constrain
the EDMs of the nuclei of $^{129}{\rm Xe}$ and $^{199}{\rm Hg}$ to be
$|d_N(^{129}{\rm Xe})|<  1.1 \times 10^{-21}\, |e|\,{\rm cm}$
and $|d_N(^{199}{\rm Hg})|< 2.8 \times 10^{-24}\, |e|\,{\rm cm}$, correspondingly.
These limits give an accuracy benchmark for proposed measurements of nuclear EDMs in ion
storage rings

This work was supported by the Australian Research Council.
We would like to thank G.~Jones for help at the initial stage of the work.
\appendix
\section{}
\label{Ap:1}
%-----------
For the matrix elements of the two operators entering \eref{Eq:U_we} (see the main text) we obtain
after certain transformations
\begin{align}
& \left\langle n^{\prime}\varkappa^{\prime}m^{\prime}\left\vert i{\bm\alpha}{\bm \sigma}\!_N
\left\{ \frac{\partial}{\partial r},\frac{r}{r_{>}^{3}}\right\}_{+}
\right\vert n\varkappa m\right\rangle \nonumber\\
&={\bm\sigma}\!_N \left\langle \varkappa^{\prime}m^{\prime}\left\vert \bf{n}\right\vert
\varkappa m\right\rangle \int_{0}^{\infty}\left[  \left(  \varkappa^{\prime
}-\varkappa-1\right)  \left(  g\frac{df^{\prime}}{dr}-f^{\prime}\frac{dg}%
{dr}\right)  \right. \nonumber\\
&+ \left.  \left(  \varkappa-\varkappa^{\prime}-1\right)  \left(  g^{\prime
}\frac{df}{dr}-f\frac{dg^{\prime}}{dr}\right)  \right]  \frac{r^{3}}{r_{>}%
^{3}}dr
\label{Ap:U1}%
\end{align}
and
\begin{align}
& \left\langle n^{\prime}\varkappa^{\prime
}m^{\prime}\left\vert \left\{ \frac{ {\bm\sigma}\!_N \mathbf{r}}{r_{>}^{3}%
}, {\bm\alpha}\mathbf{p} \right\}_{+}\right\vert n\varkappa m\right\rangle \nonumber\\
&= {\bm \sigma}\!_N \left\langle \varkappa^{\prime}m^{\prime}\left\vert \mathbf{n}\right\vert
\varkappa m\right\rangle \int_{0}^{\infty}\left[  g^{\prime}\frac{df}%
{dr}-f^{\prime}\frac{dg}{dr}+g\frac{df^{\prime}}{dr}-f\frac{dg^{\prime}}%
{dr}\right. \nonumber\\
&+ \left.  \frac{\varkappa^{\prime}+\varkappa}{r}\left(  f^{\prime
}g+fg^{\prime}\right)  \right]  \frac{r^{3}}{r_{>}^{3}}dr.
\label{Ap:U2}%
\end{align}
Adding (\ref{Ap:U1}) and (\ref{Ap:U2}), we find for the ME of the operator $U$:%
\begin{align}
& \left\langle n^{\prime}\varkappa^{\prime}m^{\prime}\left\vert U\right\vert
n\varkappa m\right\rangle =-\frac{d_N \mu_N}{Z} {\bm \sigma}\!_N \left\langle \varkappa^{\prime
}m^{\prime}\left\vert \mathbf{n}\right\vert \varkappa m\right\rangle
\nonumber\\
&  \times\int_{0}^{\infty}\left[  \left(  \varkappa^{\prime}-\varkappa\right)
\left(  g\frac{df^{\prime}}{dr}-f^{\prime}\frac{dg}{dr}-g^{\prime}\frac
{df}{dr}+f\frac{dg^{\prime}}{dr}\right)  \right. \nonumber\\
&  +\left.  \frac{\varkappa^{\prime}+\varkappa}{r}\left(  f^{\prime
}g+fg^{\prime}\right)  \right]  \frac{r^{3}}{r_{>}^{3}}dr.
\label{Eq:U_4}%
\end{align}

Using the Dirac equations for the radial wave functions
\begin{align}
\frac{df^{\prime}}{dr}  &  =-\frac{1+\varkappa^{\prime}}{r}f^{\prime}+\left(
\varepsilon^{\prime}+m+\frac{Z\alpha}{r}\right)  g^{\prime}, \nonumber \\
\frac{dg}{dr}  &  =-\frac{1-\varkappa}{r}g-\left(  \varepsilon-m+\frac
{Z\alpha}{r}\right)  f, \nonumber \\
\frac{df}{dr}  &  =-\frac{1+\varkappa}{r}f+\left(  \varepsilon+m+\frac
{Z\alpha}{r}\right)  g, \nonumber \\
\frac{dg^{\prime}}{dr}  &  =-\frac{1-\varkappa^{\prime}}{r}g^{\prime}-\left(
\varepsilon^{\prime}-m+\frac{Z\alpha}{r}\right)  f^{\prime}.
\label{Ap:De}
\end{align}
we can write
\begin{align}
&g\frac{df^{\prime}}{dr}-f^{\prime}\frac{dg}{dr}-g^{\prime}\frac{df}{dr}%
+ f\frac{dg^{\prime}}{dr} \nonumber \\
&=\frac{\varkappa^{\prime}+\varkappa}{r}\left( fg^{\prime}-f^{\prime}g\right) -
\left(  \varepsilon'-\varepsilon \right)  \left(  f^{\prime}f-g^{\prime}g\right) .
\label{Eq:gf}%
\end{align}
Now substituting (\ref{Eq:gf}) to (\ref{Eq:U_4}) we finally arrive at
\eref{Eq:U_f} given in the main text.

%\bibliographystyle{plain}
%\bibliography{./HFI_EDM}


\end{document}